\documentclass[aps,prb,reprint,superscriptaddress,longbibliography]{revtex4-1}

\usepackage{graphicx}
\usepackage{dcolumn}
\usepackage{bm}

\usepackage[utf8]{inputenc}
\usepackage[T1]{fontenc}
\usepackage{mathptmx}
\usepackage{etoolbox}

\usepackage[]{graphicx}      
\usepackage{bm}             	
\usepackage{times}			
\usepackage{tabularx}
\usepackage{color}
\usepackage{dcolumn}		
\usepackage{amsmath}
\usepackage{amssymb}
\usepackage{amsfonts}
\usepackage{calc}
\usepackage{times}
\usepackage{tabularx}
\usepackage{xcolor,colortbl}
\usepackage{rotating}

\usepackage{amsmath}

\usepackage[colorlinks,citecolor=blue,linkcolor=blue]{hyperref} 

\makeatletter
\def\@email#1#2{%
 \endgroup
 \patchcmd{\titleblock@produce}
  {\frontmatter@RRAPformat}
  {\frontmatter@RRAPformat{\produce@RRAP{*#1\href{mailto:#2}{#2}}}\frontmatter@RRAPformat}
  {}{}
}%
\makeatother
\begin{document}

\preprint{AIP/123-QED}

\title[]{
Inducing or suppressing the anisotropy in multilayers based on CoFeB}
          
\author{R. L. Seeger}
\email{rafael.lopesseeger@cea.fr.}
\affiliation{ 
SPEC, CEA, CNRS, Université Paris-Saclay, 91191 Gif-sur-Yvette, France
}
\author{F. Millo}%
\affiliation{ 
Université Paris-Saclay, CNRS, Centre de Nanosciences et de Nanotechnologies, 91120 Palaiseau, 
France
}%

\author{A. Mouhoub}%
\affiliation{ 
Université Paris-Saclay, CNRS, Centre de Nanosciences et de Nanotechnologies, 91120 Palaiseau, 
France
}%

\author{G. de Loubens}
\affiliation{ 
SPEC, CEA, CNRS, Université Paris-Saclay, 91191 Gif-sur-Yvette, France
}

\author{A. Solignac}
\affiliation{ 
SPEC, CEA, CNRS, Université Paris-Saclay, 91191 Gif-sur-Yvette, France
}

\author{ T. Devolder}
\affiliation{ 
Université Paris-Saclay, CNRS, Centre de Nanosciences et de Nanotechnologies, 91120 Palaiseau, 
France
}

\date{\today}

\begin{abstract}
Controlling the uniaxial magnetic anisotropy is of practical interest to a wide variety of applications. We study Co$_{40}$Fe$_{40}$B$_{20}$ single films grown on various crystalline orientations of LiNbO$_3$ substrates and on oxidized silicon. We identify the annealing conditions that are appropriate to induce or suppress uniaxial anisotropy. Anisotropy fields can be increased by annealing up to  11 mT when using substrates with anisotropic surfaces. They can be decreased to below 1 mT when using isotropic surfaces. In the first case, the observed increase of the anisotropy originates from the biaxial strain in the film caused by the anisotropic thermal contraction of the substrate when back at room temperature after strain relaxation during annealing. In the second case, anisotropy is progressively removed by applying successive orthogonal fields that are assumed to progressively suppress any chemical ordering within the magnetic film. The method can be applied to CoFeB/Ru/CoFeB synthetic antiferromagnets but the tuning of the anisotropy comes with a decrease of the interlayer exchange coupling and a drastic change of the exchange stiffness. 
\end{abstract}

\maketitle

\section{\label{sec:level1} Introduction}

The control of the anisotropy of a given magnetic material is very often required in the applications of magnetism \cite{spintronics2020}. 
The amorphous metallic CoFeB films are widely used in spintronics, both when very soft properties are desired such as in flux guides \cite{fluxguide}, or in contrast when a well defined uniaxial anisotropy is wanted as in the free layers of magnetoresistive field sensors \cite{spintronics2020}. 
Depending on the targeted applications, a very same material platform can even be sometimes used with opposite requirements for anisotropy. This is the case of artificial multiferroics composed of ferromagnetic films and piezoelectric layers. When meant for instance for energy harvesting, they require a well defined anisotropy \cite{Kuntal2011} while when meant for racetrack applications isotropic properties are welcome\cite{Lei2013}.
Tailoring the uniaxial anisotropy --\textit{both inducing and suppressing}-- is thus an important challenge of technological interest.


Various knobs can be employed to tune the magnetic anisotropy. Interface engineering can be used in ultrathin films\cite{handley,Chappert98}. In bulk materials one can either rely (i) on some sort of chemical ordering \cite{anisot-origin} or, (ii) on the induction of anisotropic strain in magnetostrictive materials \cite{anisoLNO}.

In the first case, one use generally saturates the magnetization using a strong magnetic field, and then provide thermal energy (hence atomic mobility) to let the structure of the material evolve towards a new state compatible with the desired magnetization orientation \cite{chikazumi}. In metallic glass like CoFeB, the anisotropy is related to some degree of alignment of the Boron atoms within the material and this can be effectively tuned and reoriented by in-field annealing \cite{anisot-origin}.
For the same reason, magnetic anisotropy can already be induced during deposition if done under an applied field \cite{depunderfield}.

The second case applies to magnetostrictive materials only. There, if an appropriate choice of the substrate influences the growth (e.g. epitaxy or strain relaxation) the resulting anisotropic strain leads to magnetic anisotropy \cite{anisoLNO}. 
This elastic coupling between the magnetic film and the substrate is systematically desired in SAW-FMR devices \cite{Weiller2011, Kuszewski2018, Rovillain2022} and magneto-acoustics \cite{Revmagnetoacoustics} when one harnesses the interaction between a surface acoustic wave (SAW) hosted by a piezoelectric substrate and the ferromagnetic resonance (FMR) of the magnetic film. 
Note that this situation fundamentally entails a dilemma when isotropic properties (meaning often stress-free layers) are desired in addition to a tight elastic coupling between film and substrate. This dilemma is significant in SAW-FMR of synthetic antiferromagnets (SAFs) since in this case a vanishing anisotropy is required for resonant coupling between the SAWs and the spin waves\cite{Verba}. Unfortunately it is difficult to obtain quasi-isotropic SAFs, one typically is left with uniaxial anisotropy fields $\mu_0 H_{\textrm{k}}$ that remain above a couple of mT \cite{CoFeB-ani1,CoFeB-ani2,CoFeB-ani3,CoFeB-ani4}.

In this paper, we study how to tailor (increase or suppress) the magnetic anisotropy of magnetostrictive layers grown on piezoelectric substrates. We develop our method on Co$_{40}$Fe$_{40}$B$_{20}$ single layer films on LiNbO$_{3}$ single crystals that are very adequate for rf acoustical waves. We show that our method is applicable to SAFs. The paper is organized as follows. We initially quantify the uniaxial anisotropy in CoFeB and show how to control it through appropriate annealing and substrate choice. The surface orientation of LiNbO$_{3}$ strongly impacts how the annealing alters the anisotropy of the magnetic material. A well designed procedure can lead to quasi-isotropic CoFeB layers, and can be extend to CoFeB/Ru/CoFeB SAFs. However, spin wave spectroscopy experiments show that the tailoring of the anisotropy of the SAF comes together with an evolution of the exchange stiffness and of the interlayer exchange coupling.

\begin{figure} 
\hspace{-1 cm}
\includegraphics[width=7cm]{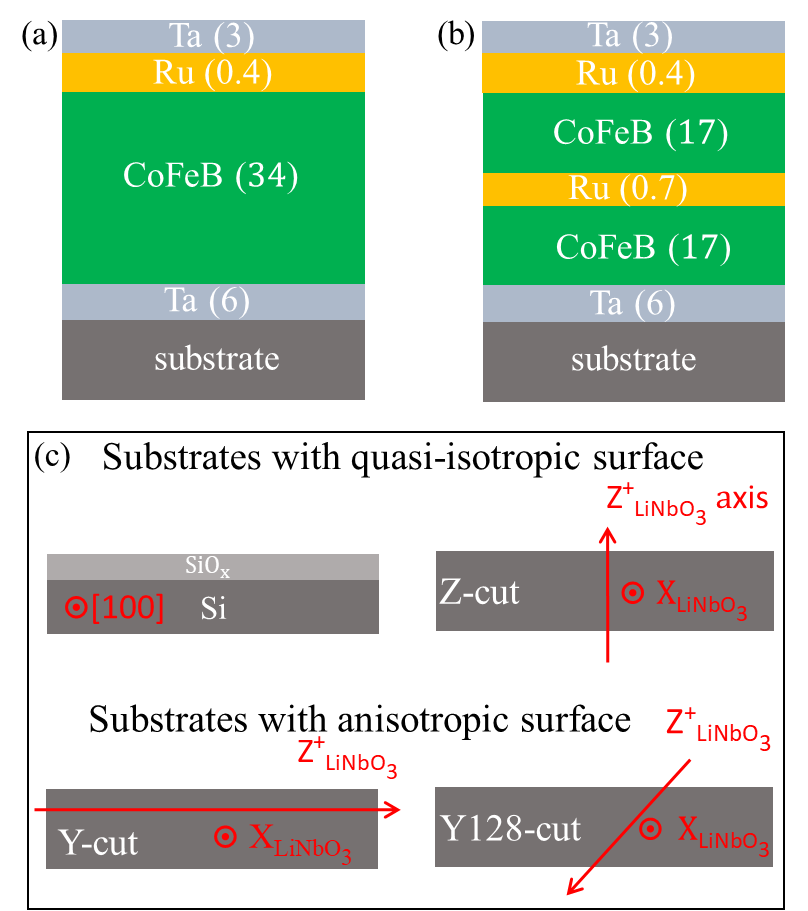}
\caption{Schematic illustration of the samples consisting of a single CoFeB layer (a) and a CoFeB/Ru/CoFeB SAF (b). All thicknesses are given in nm.  (c) Studied substrates. The red arrows indicate the direction of the ferroelectric order parameter (i.e., crystalline direction Z$_{\textrm{LiNbO}_3}^+$) around which a bulk substrate would exhibit rotational symmetry.}
\label{figStacks}
\end{figure}

 
\section{\label{sec:level2} Experiments}

\begin{figure} 
\hspace{-1 cm}
\includegraphics[width=8.5cm]{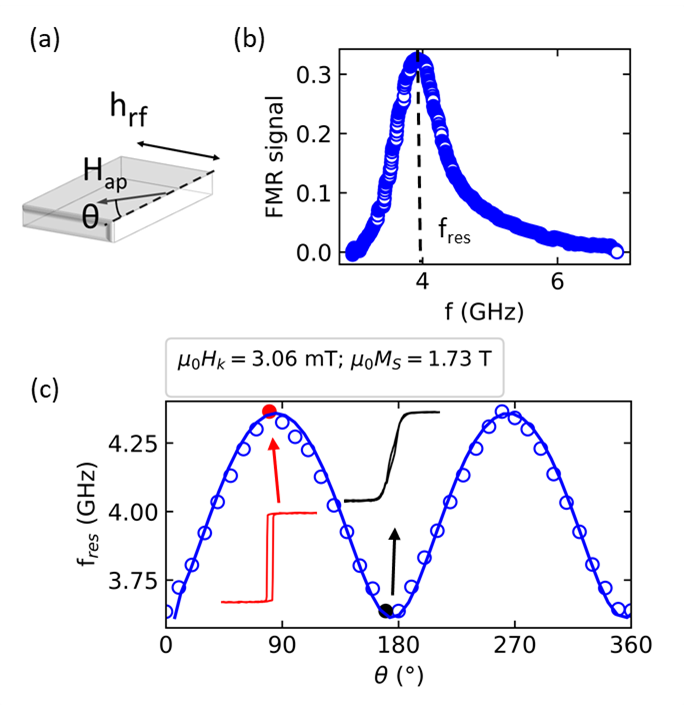}
\caption{
Schematics of the experiment and representative results for a CoFeB single film on a Y128-cut substrate in the as-grown state. (a) VNA-FMR is conducted in an in-plane orientated applied field of fixed magnitude $\mu_0 H_{\textrm{ap}}$ = 11.6 mT  and variable orientation $\theta \in [0, 360]$ deg. with respect to the long axis of the sample. (b) 
Example of VNA-FMR loss spectrum $||S_{21} (H_{\textrm{ap}})||- ||S_{21}(H=0)|| $ and definition of the resonance frequency $f_{\textrm{res}}$. (c) Symbols: experimental $\theta$-dependence of $f_{\textrm{res}}$. Line: fit within the macrospin model with uniaxial anisotropy field $H_{\textrm{k}}$ and magnetization as free parameters. Insets: experimental hysteresis loops obtained along the easy- (red curve, $\theta \cong$ 90°) and hard-axis (black curve, $\theta \cong$ 180°). 
}
\label{figexp}
\end{figure}

\subsection{Films}

Fig.~\ref{figStacks} depicts our material systems. The magnetic stacks are 
: Ta(6, buffer)/CoFeB(34)/Ru(0.4)/Ta(3, cap) (short name: single CoFeB) and Ta(6, buffer)/CoFeB(17)/Ru(0.7)/CoFeB(17)/Ru(0.4)/Ta(3, cap) (short name: CoFeB/Ru/CoFeB) SAF. All thicknesses are given in nm. The CoFeB layer was deposited from a Co$_{40}$Fe$_{40}$B$_{20}$ (at. \%) target. The deposition is done at room temperature by dc-magnetron sputtering at pressure of argon of $5 \times 10^{-3}$  mbar and base pressure below $10^{-7}$ mbar. No intentional magnetic field is applied during growth. The thickness of the Ru(0.7) spacer of the SAF is chosen to maximize the interlayer exchange coupling \cite{Asma2022}. 

\subsection{Substrates}

The depositions were done on several substrates ranging from naturally oxidized silicon wafers (to be referred as Si/SiO$_{\textrm{x}}$) to LiNbO$_{3}$ single crystals of various surface orientation (Z-, Y- and Y128-cut) \cite{LiNbO3}. Since the properties of the magnetic materials can be impacted by the stress induced by the underlying substrate\cite{LNOexpansion}, we cast the substrates in two categories. The first category gathers the substrates whose surface expands in a quasi-isotropic manner upon annealing. For the second category the thermal expansion is anisotropic at the surface, as illustrated in Fig. \ref{figStacks}(c). 

\subsection{Post-growth annealing conditions}

In order to unravel the respective roles of substrate induced applied stress, applied field, and Boron diffusion onto the annealing-induced evolution of the magnetic properties, we have annealed our material systems using 4 different procedures: \begin{itemize}
\item  without any applied magnetic field.
\item  with a 70 mT applied in a given direction in a 1 step manner. We will see that the field direction (i.e., along or orthogonal to the initial anisotropy axis) shall not influence the final result.
\item  With a 70 mT field applied in 2 successive steps: the sample is first annealed with a field oriented at some randomly chosen in-plane direction, then at along an orthogonal direction.
\item  In a 30 mT field rotating at 5 rpm in the sample plane.
\end{itemize}
The annealing temperature $T$ ranges from 100 to 200 °C, above which systematic crystallization is expected for our Boron content \cite{you2008,conca2014,Thibaut2018,sriram2023}. The annealing time is set to 4 min on a hot plate for (i), (ii) and (iii). The annealing in rotating field (iv) is done in vacuum for 10 h. In all cases the field is applied while ramping up and down the temperature and is strong enough to saturate the magnetization. 

\subsection{Magnetic characterizations}

The  magnetic characterization of samples was performed by vibrating sample magnetometry (VSM) and Vector Network Analyzer ferromagnetic resonance \cite{vnafmr} (VNA-FMR). An in-plane applied field $\mu_0 H_{\textrm{ap}}$ was used, and its direction $\theta$ [see Fig. \ref{figexp}(a)] was varied to access to the sample's magnetic anisotropy. The resonance spectra (FMR absorption signal) are obtained by measuring the field dependence of the VNA transmission parameter $||S_{21} (H_{\textrm{ap}})||- ||S_{21}(H=0)||$, as plotted in Fig. \ref{figexp}(b). The resonance frequencies (FMR for the single CoFeB or acoustical and optical resonances of the SAF) $f_{\textrm{res}}$ are defined from the maxima of absorption.

The $\theta$-dependence of the FMR of single CoFeB films were analyzed in the macrospin approximation using numerical energy minimization and subsequent application the Smit-Beljers equation\cite{SmitBeljers}. A fitting procedure allowed to extract independently the values of the uniaxial anisotropy field $H_\textrm{k}$, the orientation of the easy axis and the saturation magnetization $M_\textrm{s}$. Fig.~\ref{figexp}(c) illustrates this procedure when applied to a single CoFeB film grown on a Y128-cut substrate in the as-grown state. 
The orientation of the easy axis and the uniaxial character of the anisotropy are systematically consistent with the hysteresis loops. 


The  $\theta$ and $H$ dependence of the acoustical $f_\textrm{acou}$ and optical $f_\textrm{opt}$ resonances of the SAF were analyzed in the full micromagnetic framework \cite{Vansteenkiste2014} following the method described in ref.~\onlinecite{Asma2022}. There, it was shown that the competition between the interlayer coupling $J$ and the intralayer exchange stiffness $A_\textrm{ex}$ results in the existence of a gradient of the magnetization orientation in the growth direction. This gradient renders the curvature of the $f_\textrm{opt} (\theta)$ near $H=0$ very sensitive to the ratio of $A_\textrm{ex}$ and $J$, that can thus be deduced reliably. A fitting procedure of $f_\textrm{acou}(\theta)$ can then be used to extract the anisotropy fields, assumed to be exactly the same for the two magnetic layers of the SAF. 

\section{\label{sec:level3} Evolution of the magnetic anisotropy upon annealing} 

\subsection{Results} 

\begin{figure} 
\hspace{-1 cm}
\includegraphics[width=8cm]{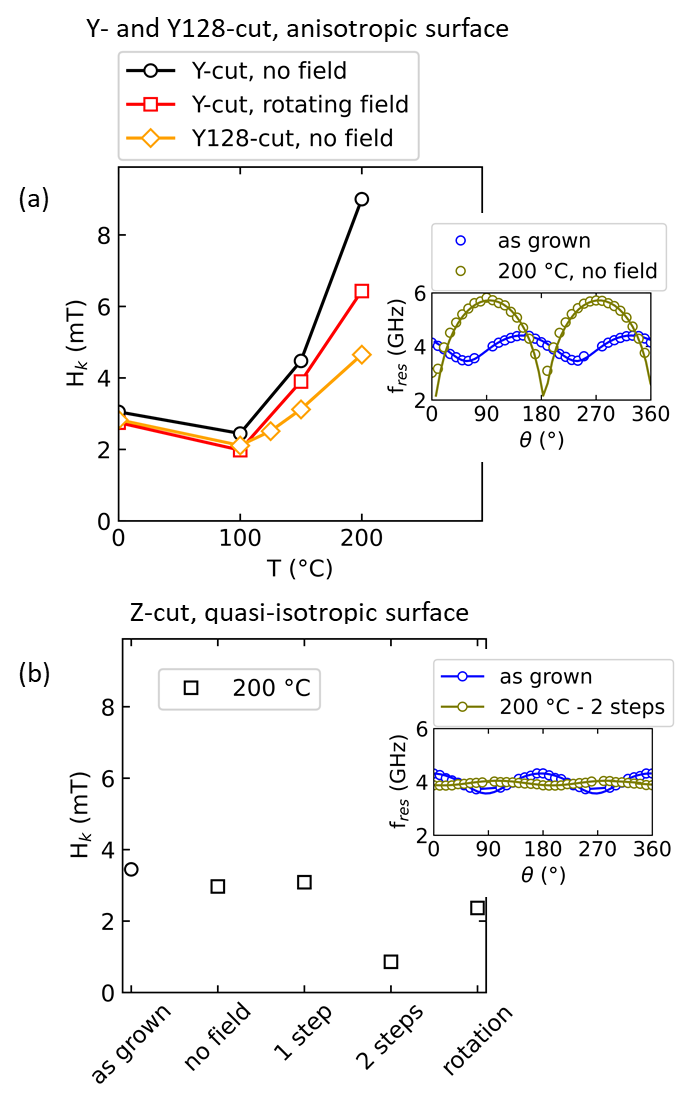}
\caption{(a) Anisotropy field $H_{\textrm{k}}$ dependence with the annealing temperature $T$ as measured for single CoFeB grown on Y-cut LiNbO$_3$, as an example of effect of annealing on a substrate with anisotropic surface. (b) Representative $H_{\textrm{k}}$-dependence for samples subjected to various annealing procedures with $T$ = 200 °C , as measured for single CoFeB grown on Z-cut LiNbO$_3$, a substrate with quasi-isotropic surface. Insets in (a) and (b) are representative $\theta$-dependence of $f_{\textrm{res}}$ before and after annealing. The line is a fit to the experimental data. 
}
\label{figproc}
\end{figure}

 The main features of the evolution of the magnetic anisotropy upon annealing are illustrated in Fig.~\ref{figproc} and compiled in Table~\ref{tableCoFeB}. An annealing temperature above 100°C appeared necessary to observe an evolution of the magnetic properties. The atomic mobility within the CoFeB films is likely insufficient below this temperature. For larger annealing temperatures, the magnetic anisotropy evolves in very different ways depending if the substrate has an isotropic or an anisotropic surface, and also on the field applied.
 
 When working on substrates with anisotropic surfaces (Y-cut and Y128-cut LiNbO$_3$), the annealing increases substantially the anisotropy, see Fig.~\ref{figproc}(a). 
The inset compares for instance the angular dependence of the FMR of a CoFeB film on a Y-cut LiNbO$_3$ substrate before and after a 200 °C field-free annealing. Annealing increases the anisotropy field $\mu_0H_{\textrm{k}}$ from 3.05 mT to 9.00 mT. This comes with a reorientation of the hard axis towards the X$_{\textrm{LiNbO}_3}$ (i.e., $\theta$ = 0 deg. in our convention). 
The Y128-cut samples follow a similar trend but with a lower increase of the anisotropy. As soon as the anisotropy increases, the hard axis also reorients towards X$_{\textrm{LiNbO}_3}$ axis and the easy axis towards the in-plane projection of the Z$_{\textrm{LiNbO}_3}$ axis.
When working on these substrates with anisotropic surfaces, the magnetic field applied (or not) during the annealing has a minor influence on the evolution of the magnetic anisotropy. 

Conversely, the anisotropy can be reduced for the films grown on the substrates with quasi-isotropic surfaces: Z-cut LiNbO$_3$ and oxidized silicon. The inset in  Fig.~\ref{figproc}(b) shows for instance angular dependence of the FMR before and after a 2 steps annealing procedure in the case of a Z-cut LiNbO$_3$ substrate. This 2-step annealing lowers the anisotropy down to 0.6 mT. Notably, the rate of decrease of the anisotropy depends strongly on the field sequence used during annealing, and the hard axis systematically ends perpendicular to the field applied during the last annealing step. 
 
\subsection{Physical origins of the evolution of anisotropy} 

The previous results can be discussed by considering two thermodynamic phenomena: (i) the interplay between magneto-elasticity and anisotropic strain and, (ii) the chemical ordering within the magnetic material. We recall that for the annealing temperatures studied here, no crystallization of the CoFeB layer is expected. Let us first discuss the magneto-elastic scenario.

\begin{table} 
\caption{Summary of the material parameters in single CoFeB subjected to different thermal treatments. "Variable" means that the value of the anisotropy field and its orientation depend on the sample position within the deposition machine -- the evolution of $H_k$ is however taken for a consistent sample position.  \\ 
}
\label{tableCoFeB}
\begin{tabular}{| c | c | c | c |  c |}
 \hline
 Sub. & Thermal & $\mu_0 H_k$ (mT) & Hard axis & $\mu_0 M_s$ (T)   \\ 
  & treatment & $\pm 5 \% $ & & $\pm 0.01$    \\ \hline
  \multicolumn{4}{l}{Substrates with quasi-isotropic surface:}\\ \hline
SiOx & as grown  & 3.0 & variable & 1.70   \\ 
 & 200$^\circ$C, H=0 & 3.1 & variable & 1.72     \\ 
 & 200$^\circ$C, rotH &  1.5 & variable & 1.71     \\ 
 & 200$^\circ$C 2-step H & 0.9 & $\perp$ to last field  & 1.76     \\ \hline
Z & as grown  & 2.2 & variable & 1.70  \\ 
 & 200$^\circ$C, H=0  & 2.8 & variable & 1.71     \\ 
 & 200$^\circ$C, rotH  &  2.4 & variable &  1.82    \\ 
 & 200$^\circ$C 2-step H  & 0.6 & $\perp$ to last field  &  1.70    \\ \hline 
  \multicolumn{4}{l}{Substrates with anisotropic surface:}\\ \hline
Y128 & as grown  & 3.1 & variable &  1.77 \\ 
 & 200$^\circ$C, H=0 & 4.3 & $\parallel$ X$_{\textrm{LiNbO}_3}$  & 1.87     \\ 
 \hline
Y & as grown  & 3.0 & variable & 1.70  \\ 
 & 200$^\circ$C, H=0  & 9.0 &  $\parallel$ X$_{\textrm{LiNbO}_3}$ & 2.00    \\ 
 & 200$^\circ$C, rotH  & 6.4 & $\parallel$ X$_{\textrm{LiNbO}_3}$ & 2.00     \\
& 200$^\circ$C 2-stepH  & 11.4 & $\parallel$ X$_{\textrm{LiNbO}_3}$ & 2.00     \\ \hline
\end{tabular}
\end{table}

\subsubsection{Magneto-elastic scenario} 

When annealing above $T_{\textrm{a}}>$ 100°C, the atoms within the CoFeB film acquire some mobility, as observed in other soft magnetic materials \cite{anisot-origin}. Being amorphous, the glassy CoFeB film slowly flows, such that after a sufficient delay it reaches a relaxed (stress-free) state at the annealing temperature.  
Cooling (defining $RT- T_{\textrm{a}} = \delta T <0$) to room temperature (RT) quenches suddenly any atomic mobility, while triggering a thermal contraction. Since the CoFeB film is clamped by the much thicker substrate, the in-plane strain $\epsilon$ of the substrate is imposed to the magnetic film. The natural contraction of an hypothetically free-standing CoFeB film would be isotropic (its thermal expansion coefficient $\beta$ is isotropic). That of the LiNbO$_3$ substrate is not: the thermal expansion coefficient in the X$_{\textrm{LiNbO}_3}$ direction is stronger than in the other direction of the substrate plane \cite{LNOexpansion}. As a result the stress $\bar {\bar {\epsilon}}$ within the CoFeB at RT is biaxial and more compressive in the X$_{\textrm{LiNbO}_3}$ direction. Defining $x$ and $y$ as the two directions of the surface of the substrate with $x \parallel$ X$_{\textrm{LiNbO}_3}$ [see Fig. 1(c)], we have:
\begin{equation}
    \epsilon_{xx}= \beta_x \delta T  <0 \textrm{~and~} \epsilon_{yy} = \beta_y \delta T <0
\end{equation}
Whatever the substrate cut, the largest deformation is always along X$_{\textrm{LiNbO}_3}$, see ref.~\onlinecite{LNOexpansion}. We have $\beta_x > \beta_y$ for both the Y-cut case and the Y128-cut case. 
As the CoFeB has essentially a free surface, its stress is purely biaxial, such that there is no shear strain (i.e., $\epsilon_{xy} =0$).

This biaxial strain generates a magnetic anisotropy of CoFeB. Indeed for an in-plane magnetized film, the magneto-elastic energy is
$    E_{\textrm{me}}=B_1  (m_x^2 \epsilon_{xx}+ m_y^2\epsilon_{yy}) + B_2 m_x m_y \epsilon_{xy} $, where $B_1$ and $B_2$ are the usual magneto-elastic coefficients. In amorphous materials, they reduce to $-\frac{3} {2} \lambda E_{\textrm{Young}}$ which amounts to $ -7.6$~MJ/m$^3$ with the magnetostriction coefficient $\lambda=27~\textrm{ppm}$ for Co$_{40}$Fe$_{40}$B$_{20}$ from ref.~\onlinecite{magnetostriction-CFB} and the Young's modulus $E_{\textrm{Young}} = 187$ GPa, from ref.~\onlinecite{Peng2016}. Note that $\lambda >0$, meaning a tensile strain lowers the energy. The CoFeB film being more compressed in the X-direction than in other directions, X$_{\textrm{LiNbO}_3}$ will become the hard axis.
Using the conservation of the magnetization norm and $\epsilon_{xy} =0$, we can rewrite this energy in the form of an effective uniaxial anisotropy: 
\begin{equation} E_{me}=B_1 m_x^2 (\epsilon_{xx}-\epsilon_{yy})
\end{equation}
with a magneto-elastic effective anisotropy field of: 
\begin{equation}  \mu_0 H_k^\textrm{mel}=\frac{2 B_1}{M_s} (\beta_x-\beta_y) \delta T \label{melHk} \end{equation} that is predicted to be linear with the annealing temperature, which bears some similarity with the experimental results [see Fig.~3(a)] above $T_{\textrm{a}}$=100°C.

\begin{table} 
\caption{Thermal expansion coefficients in the two directions of the surface plane, in units of $10^{-5}/^\circ$C. \\ 
}

\label{tableLNO}
\begin{tabular}{| c | c | c | c |}
 \hline
     
     Surface & $\beta$ in first direction & $\beta$ along a perp  direction \\ \hline
Y-cut  & along X$_{\textrm{LiNbO}_3}$: $\beta_{1}=1.5$  &  along X$_{\textrm{LiNbO}_3}$: $\beta_3=0.7$    \\ \hline

Y128-cut & along X$_{\textrm{LiNbO}_3}$:  $\beta_{1}=1.5$ & $\beta_{1} \sin^2(128)+ \beta_{3} \cos^2(128)=1.2$   \\ \hline
Z-cut & along X$_{\textrm{LiNbO}_3}$:  $\beta_{1}=1.5$  & along X$_{\textrm{LiNbO}_3}$: $\beta_{1}=1.5$   \\ \hline
Si[001] & along [100]: 0.468  & along [010]: 0.468   \\ \hline
 
\end{tabular}
\label{expLNO}
\end{table}

Using the data of Table~\ref{tableLNO} with $\delta T = -180$ K, Eq.~\ref{melHk} predicts $\mu_0 H_k^\textrm{mel}= 5.5\textrm{~mT}$ for the Y128-cut case and $\mu_0 H_k^\textrm{mel}= 13.7\textrm{~mT}$ for the Y-cut case, with hard axes along X$_{\textrm{LiNbO}_3}$ in both cases. This magneto-elastic contribution dominates any other contribution to the uniaxial anisotropy, including the one present in the as-grown state and those possibly related to the magnetic field applied during the annealing. This correlates with our finding on the minor influence of the field applied during the annealing of Y-cut and Y128 samples 
Note that the predicted values of the anisotropy field $H_k^\textrm{mel}$ are slightly marger 
than our experimental findings. This may indicate that the stress release is incomplete during the annealing, or that the magnetostriction coefficient of the literature \cite{magnetostriction-CFB} is over estimated. 

Another conclusion of our study concerns the evolution of the magnetization $M_{\textrm{s}}$ upon annealing (Table \ref{tableCoFeB}). There is little evolution for the Si/SiO$_{\textrm{x}}$ case, but a substantial increase otherwise. For our nominal Boron concentration, bulk crystallisation or stress-induced bulk crystallization are not supposed to occur at our annealing temperatures \cite{trexler_mechanical_2010} and can therefore not be invoked for the observed increase of the magnetization. However, some compression-induced diffusion of the Boron atoms out of the magnetic films may start to occur, thereby reducing the Boron concentration. The corresponding densification is generally associated with an increase of the magnetization \cite{kim_control_2022}. We indeed observe that the increase of the magnetization after the 200°C annealings seems to correlate with the amount of compression (see Tables \ref{tableCoFeB} and \ref{tableLNO}).

This anisotropic strain-induced scenario is effective only for the substrates with anisotropic thermal expansion like Y-cut and Y128-cut LiNbO$_3$. Another scenario must thus be invoked for the Z-cut of LiNbO$_3$ and the oxidized silicon cases. 

\subsubsection{Chemical order scenario} 


In many magnetic alloys, it is routinely observed that annealing in a magnetic field induces a preferred direction of magnetization \cite{handley,chikazumi}. A plausible model often invoked to explain this mechanism is the migration of atoms on a local scale in such a way as to favor magnetization in a given direction. At annealing temperatures leading to some atomic mobility, some atom pairs orient themselves relative to the direction of magnetization set by the field, so as to decrease their magnetic anisotropy energy. Cooling to a temperature where atomic diffusion gets quenched, the anisotropy axis remains along the direction it has acquired during annealing. Metalloids like Boron play an important role in this process thanks to their high mobility and chemical interaction with transition metals \cite{anisot-origin}. In metallic glass like CoFeB the applied field drives an anisotropic distribution of atoms pairs among Co, Fe and B, and this mechanism is active at the time scales used for annealing for the temperatures considered here \cite{anisot-origin}. 
This process is generally used to induce anisotropy, using a \textit{fixed} field orientation and a \textit{long} annealing.

However it is important to figure out that the material evolution is a thermodynamical process: during our first annealing step, the field-induced (energy-minimization-driven) chemical ordering process competes with a temperature-induced (random, entropy-driven) disordering trend. This competition leads to a slow formation of a uniaxial anisotropy.
During the subsequent annealing step, the magnetic field orientation is different. As a result, the entropy-driven and energy-driven thermodynamical forces both tend to destroy the previously favored chemical order, and therefore they concur to reduce the anisotropy. Because of this coincidence of the two thermodynamical forces at play, this destruction of the previously set anisotropy is a fast process. The building-up of any uniaxial anisotropy along a new field direction is a much slower process. In practice, it is not seen at the time scales used in our annealings.

This explains how one can progressively reduce the anisotropy by applying successive orthogonal fields during annealing when there is no magneto-elastic contribution at play. Each annealing step with a new field direction statistically breaks pair alignments which results in a progressive randomization of the chemical ordering within the magnetic film and, thus, a decrease of the anisotropy. 
The effect of annealing under a rotating field is qualitatively similar as a two-step annealing, but their relative efficiency depends on the characteristic time-scales of atomic diffusion versus the field rotation period.

To summarize the discussion so far, the findings presented in Fig. \ref{figproc} indicate that, with regards to the substrate, different contributions to the uniaxial anisotropy of CoFeB may arise. For surface substrates with anisotropic thermal expansion (Y-cut and Y128-cut LiNbO$_3$), the anisotropy is controlled by anisotropic strain, while for quasi-isotropic surface substrates (Si/SiO$_{\textrm{x}}$ and Z-cut LiNbO$_3$), the anisotropy is controlled by the chemical ordering favored or broken by the sequence of applied magnetic fields.

\section{\label{sec:level4} Applicability to synthetic antiferromagnets}

\begin{figure} 
\hspace{-6mm}
\includegraphics[width=8.5cm]{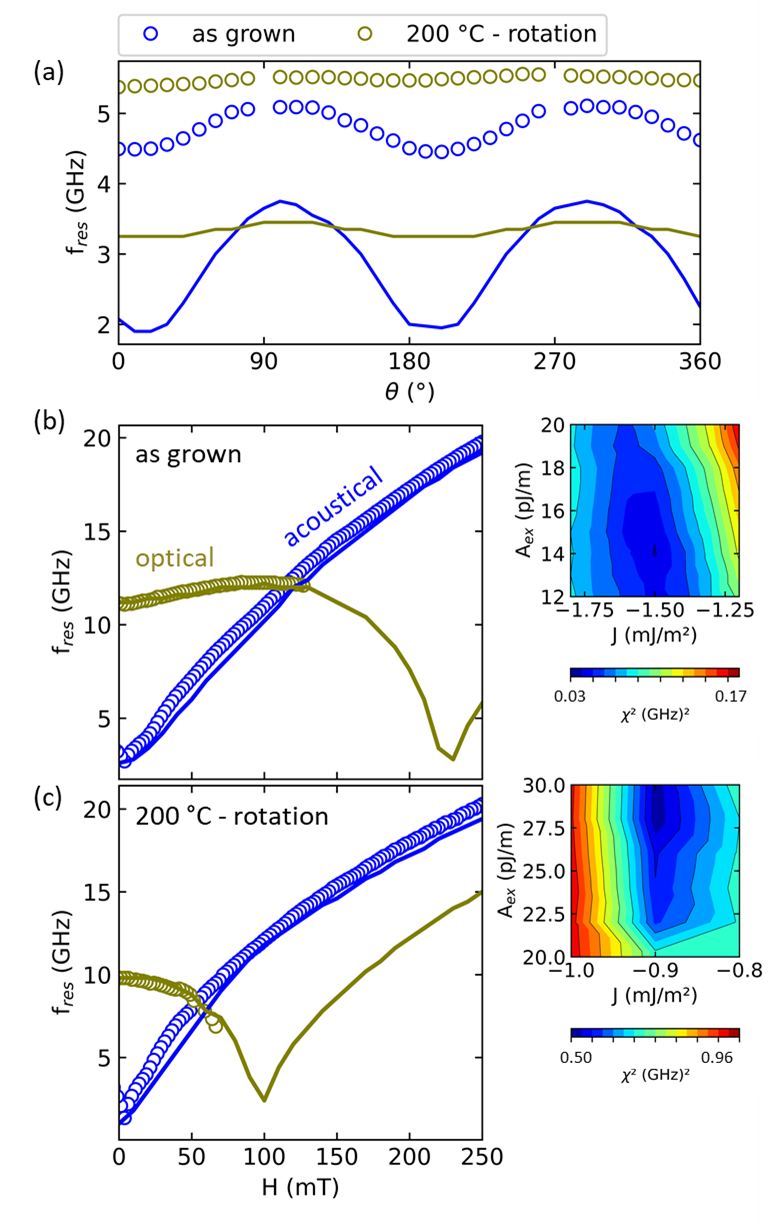}
\caption{Effect of annealing in a rotating field for CoFeB/Ru/CoFeB SAFs grown on Si/SiO$_x$. (a) $\theta$-dependence of $f_{\textrm{res}}$ before and after annealing. Symbols are experimental data. The lines show the calculated dependencies from micromagnetic simulations \cite{Vansteenkiste2014} using magnetic parameters fitted from the broadband VNA-FMR characterization of the acoustical and optical modes shown in (b) in the as-grown state and in (c) after annealing. Insets show color maps of the distance $\chi$ between the experimental and simulated spin wave frequencies used to determine $A_{\textrm{ex}}$ and $J$.
}
 
\label{figsimul}
\end{figure}

It is important to investigate whether the conclusions previously established for single CoFeB films can be extended to multilayers. In particular, let us see if one can obtain isotropic CoFeB/Ru/CoFeB synthetic antiferromagnets (SAF) when grown on quasi-isotropic substrates. Fig. \ref{figsimul} shows the results for a SAF grown on Si/SiO$_{\textrm{x}}$ in the as-grown state and after annealing in a rotating field. The two lowest order spin wave modes can be detected, the acoustical ($f_{\textrm{acou}}$) and the optical ($f_{\text{opt}}$) modes. The value of $f_{\text{acou}}$ at low field is known to be very sensitive on the anisotropy field\cite{DevolderSAF}. The $\theta$-dependence of $f_{\textrm{acou}}$ [Fig.\ref{figsimul}(a)] clearly indicates that the annealing procedure defined for the single layer films succeeded also in suppressing the anisotropy of the SAF grown on Si/SiO$_{\textrm{x}}$. As shall be explained later, we can only give a semi-quantitative measurement the anisotropy field $\mu_0H_{\textrm{k}}$ of the SAF, but its reduction (see Table \ref{tableSAF}) is almost complete. The same trend is observed when the growth is performed on Z-cut LiNbO$_3$ (see Table \ref{tableSAF}). 

However this quasi-suppression of the anisotropy is accompanied by an evolution of the other magnetic properties.  This can be seen by comparing the frequencies of the experimental and simulated modes using the methodology defined in ref.~\onlinecite{Asma2022}. 
Indeed the value of $f_{\text{opt}}$ at $H=0$ is essentially set by the interlayer coupling $J$.     Annealing obviously reduces it [compare Fig.~\ref{figsimul}(b) and (c)]. Besides, the curvature of $f_{\text{opt}}$ versus $H$ is very sensitive to the ratio $\frac{A_{\textrm{ex}}}{J}$: annealing obviously strongly affects this ratio.

The values of $A_{\textrm{ex}}$,  $J$ that best fit the experimental data for Si/SiO$_{\textrm{x}}$ and Z-cut LiNbO$_3$ substrates are listed in Table \ref{propSAF}. The experiment-to-micromagnetics agreement is excellent except in the small field region for the acoustical spin wave, where micromagnetics systematically underestimates the frequency of the acoustical mode [Fig.~\ref{figsimul}(b) and (c)]. The same difficulty arises when attempting to account for the $\theta$-dependence of $f_{\textrm{acou}}$ with micromagnetic simulations, as shown in Fig.~\ref{figsimul}(a).   
The reason for this disagreement was not identified, but we believe that it may arise from a gradient of the magnetic properties in the growth direction which is not taken into account in the simulations.
For this reason, we can only give a semi-quantitative measurement the anisotropy field $\mu_0H_{\textrm{k}}$. The anisotropy values in Table \ref{propSAF} are deduced from the sole value of $f_{\text{acou}}$ at zero field. 

\begin{table} 
\caption{Material parameters of the synthetic antiferromagnet before and after annealing in a rotating field. 
}
\label{tableSAF}
\begin{tabular}{| c | c | c | c | c | c |}
 \hline
 Substrate   & $\mu_0 H_k$ (mT) & $\mu_0 M_s$ (T) & $J$ (mJ/m$^{2}$) & $A_{ex}$ (pJ/m)   \\ \hline
SiO$_{x}$, as grown  & 4.4
& 1.70  & -1.5 & 14.5  \\ \hline

SiO$_{x}$, 200$^\circ$C &  0.8  & 1.71  & -0.9 & 28.3   \\ \hline

Z$_{\textrm{LiNbO}_{3}}$, as grown  & 3.8 & 1.70 & -1.7 & 15.5 \\ \hline

Z$_{\textrm{LiNbO}_{3}}$, 200$^\circ$C  & 0.5 & 1.70 & -1.1 & 21   \\ \hline

\end{tabular}
\label{propSAF}
\end{table}

Upon annealing, $J$ reduces from -1.5 mJ/m$^2$ to -0.9 mJ/m$^2$ upon annealing, as observed for our SAF grown on Si/SiO$_{\textrm{x}}$. The atomic mobility enabled by the annealing probably reduces the sharpness of the interfaces of the Ru spacer, thereby decreasing $J$. 
The evolution of the local order within the CoFeB material is also evident from the evolution of its exchange stiffness $A_{\textrm{ex}}$, which undergoes a very substantial increase from 14.5 pJ/m to 28.3 pJ/m upon annealing. It is noteworthy that the as-grown value of $A_{\textrm{ex}}$ is comparable to literature values in the amorphous state for our composition, which are found to be \cite{Asma2022,A-cofeb1,A-cofeb2} in the range from 10 to 14 pJ/m. The exchange stiffness is also known to increase substantially when the layer get either crystalline or simply more dense \cite{kim_control_2022}.

\section{Conclusion}

We have studied the impact of annealing on the magnetic properties of CoFeB films and synthetic antiferromagnets. We identified the different contributions to the uniaxial anisotropy in CoFeB single films by performing various in-field thermal treatments for films grown on different substrates. The anisotropy field of CoFeB can be increased  when the annealing is performed in samples grown on substrates whose surfaces have an anisotropic thermal expansion. In this case the likely scenario is a full stress relaxation occurring during the annealing, followed by the creation of a biaxial strain in the CoFeB upon cooling, which induces a strong magneto-elastic anisotropy. Anisotropy fields up to 11 mT can be induced when extremely anisotropic substrates like Y-cut LiNbO$_3$ are used. Conversely, the anisotropy field can be decreased to below 1 mT when using substrates whose surface is quasi-isotropic. In this case the anisotropy is controlled by the history of the magnetic field applied during annealing. In particular, sequences of orthogonal fields are very efficient in suppressing the anisotropy. This method was applied to obtain isotropic CoFeB/Ru/CoFeB synthetic antiferromagnets; however the annealing also affects the exchange interactions within the stack.

\begin{acknowledgments}
We acknowledge discussions with S. Margueron and A. Bartasyte. This work was supported by a public grant overseen by the French National Research Agency (ANR) as part of the “Investissements d'Avenir” program (Labex NanoSaclay, reference: ANR-10-LABX-0035, project SPICY). R. L. S and F. M. acknowledge the French National Research Agency (ANR) under Contract No. ANR-20-CE24-0025 (MAXSAW).

\end{acknowledgments}





\section*{References}

\bibliography{aipsamp}

\end{document}